\documentclass[letter]{aa} 

\usepackage{graphicx}
\usepackage{txfonts}
\begin{document}

\title{The eccentric short-period orbit of the supergiant fast X-ray transient HD~74194 (=LM Vel)}


   \author{
R. Gamen\inst{1}
\and
R. H. Barb\'a\inst{2} 
\and
N. R. Walborn\inst{3}
\and
N. I. Morrell\inst{4}
\and
J. I. Arias\inst{2}
\and
J. Ma\'iz Apell\'aniz\inst{5}
\and
A. Sota\inst{6}
\and
E. J. Alfaro\inst{6}
          }

\institute{Instituto de Astrof\'{\i}sica de La Plata, CCT La Plata-CONICET,
Facultad de Ciencias Astron\'omicas y Geof\'{\i}sicas, Universidad Nacional de
La Plata, Paseo del Bosque S/N, (1900) La Plata, Argentina.\\
              \email{rgamen@fcaglp.unlp.edu.ar}
        \and
 Departamento de F\'{\i}sica, Universidad de La Serena, Av. Juan Cisternas 1200
 Norte, La Serena, Chile. 
\and
Space Telescope Science Institute\thanks{Operated by AURA, Inc., under
  NASA contract NAS5-2655}, 3700 San Martin Drive, Baltimore, MD 21218, USA
\and
 Las Campanas Observatory, Carnegie Observatories, Casilla 601, La Serena,
 Chile. 
\and
Centro de Astrobiolog\'ia, CSIC-INTA, campus ESAC, apartado postal 78, E-28 691 Villanueva de la Ca\~nada, 
Madrid, Spain.
\and
Instituto de Astrof\'{\i}sica de Andaluc\'{\i}a-CSIC, Glorieta de la
Astronom\'{\i}a s/n, 18008, Granada, Spain.            
}

   \date{}

 
  \abstract
   {}
   {We present the first orbital solution for the O-type supergiant star
     HD~74194, which is the optical counterpart of the supergiant fast X-ray
     transient IGR~J08408-4503.} 
   {We measured the radial velocities in the optical spectrum of HD~74194, and we
     determined the orbital solution for the first time. We also analysed the complex H$\alpha$ profile.}
   {HD~74194 is a binary system composed of an O-type supergiant and a compact object in a short-period
   ($P$=9.5436$\pm$0.0002 d) and high-eccentricity ($e$=0.63$\pm$0.03) orbit. 
     The equivalent width of the H$\alpha$ line is not modulated entirely with the orbital period, 
     but seems to vary in a superorbital period ($P$=285$\pm$10 d) nearly 30 times longer than the orbital one.}
   {}

   \keywords{Stars: binaries: spectroscopic --
                stars: early-type -- 
                X-rays: individual (IGR J08408--4503)
                stars: individual (HD 74194)
               }

   \maketitle
%

\section{Introduction}

HD~74194 (=LM~Vel) is a supergiant O-type star 
\citep[O8.5~Ib-II(f)p;][]{2014ApJS..211...10S}\footnote{HD~74194 was first identified as a
supergiant by \citet{1973AJ.....78.1067W}, who classified it as O8.5~Ib(f).}.
It was first suspected of binarity by \citet{1987ApJS...64..545G}, based on
the two radial-velocity (RV) measurements published by \citet[][-4.3 km
s$^{-1}$]{1955MmRAS..67...51F} and \citet[][22.4 km
s$^{-1}$]{1977ApJ...214..759C}. No further analysis was found in the
literature.  

Because the binary nature of HD~74194 was uncertain, we included it in the
sample of a RV monitoring programme. This project, named the {\em OWN Survey},
is carrying out systematic spectroscopic observations of all the O- and WN-type
stars known in our Galaxy \citep{2010RMxAC..38...30B}, for which there was no
or weak indication of binarity or multiplicity. 
The major goals are to search for RV variations that are indicative of
orbital motion and to derive orbital solutions for as many of the variables as possible.
The project also aims at building a high-quality spectral library of O-type stars 
to be used as a first epoch of RVs for future searches and also as spectral variability studies.
The current sample of the {\em OWN Survey} includes the southern O-type stars from
version 1.0 of the Galactic O-Star Catalog \citep{gosc} and some Wolf-Rayet stars of the nitrogen 
sequence (WN) from the 7th Catalog of Galactic WR Stars \citep{2001NewAR..45..135V}.
The {\em OWN Survey} started in 2005, and it has collected more than 6000 high-resolution spectrograms to date.

\citet{2006ATel..813....1G} report the discovery of an {\it INTEGRAL} source, IGR J08408--4503, which
they associated with HD~74194, although they could not discard the possibility of it being 
a very long and soft gamma-ray burst (GRB) at high redshift.
Almost immediately, \citet{2006ATel..814....1M} found a previous, unreported outburst in the 
{\it INTEGRAL} archive, making the GRB hypothesis highly unlikely. They identified the source as a
supergiant fast X-ray transient (SFXT), just because of the presence of the supergiant HD~74194.
A similar conclusion was reached by \citet{2006ATel..815....1M}. They analysed ESO optical archival spectra, 
and the presence of H$\alpha$ in emission justified the optical--X-ray connection. 
At that time, the {\em OWN Survey} was beginning, and a few optical spectra were available. 
We analysed them and detected variability in both the H$\alpha$ profiles and the RVs,
which pointed to a binary nature.
We therefore reported observational evidence to interpret HD~74194 as a SFXT object 
\citep{2006ATel..819....1B}.
   
We continued observing the optical spectrum of HD~74194 within the {\em OWN Survey} programme until this year, 
and we found a definitive orbital solution.
Although several studies were published following the initial rapid sequence of telegrams 
(between 18 and 23 May 2006) analysing the hard X-ray source as a SFXT, this paper provides the first  
direct proof that HD~74194 is a binary system and thus that the identification is correct.

SFXTs are supposed to be binary systems comprised of a blue supergiant star and a compact 
(usually neutron-star) object 
displaying extreme transient flaring activity in the X-ray domain
(see the review of \citealt{2013arXiv1301.7574S} for details).
This more extreme X-ray variability is the criterion proposed to distinguish SFXTs from the 
`classical' supergiant X-ray binaries, which share other macroscopic properties such as the range of
orbital periods \citep{2015AdSpR..55.1255B}.
The mystery of SFXTs is the origin of this difference.
To date, there are only twelve confirmed SFXTs
listed in \citet{2014A&A...562A...2R} and \citet{2015JHEAp...7..126R}.
IGR~J08408--4503~=~HD~74194 is one of the three that display the broadest X-ray dynamic ranges ($\geq 10^4-10^5$) and
that are indicated as an ``SFXT prototype'' \citep{2015AdSpR..55.1255B}.

In this work, we present the optical spectroscopic observations and the RV orbit of HD~74194. 
We also study the coincidence of the reported flares
with the periastron passages, which strongly suggests a link between the two events.
In addition, we study the variability of the H$\alpha$ profile, searching for a superorbital period.


\section{Observations}

This work is based on spectroscopic observations with high-resolution spectrographs at Complejo 
Astron\'omico El Leoncito (CASLEO) in Argentina, 
and at the Las Campanas (LCO) and La Silla (ESO) Observatories in Chile.
A summary of the spectral characteristics is provided in Table~\ref{runs}.

We obtained a calibration lamp exposure immediately before or after each target integration in the same
telescope position when the spectrograph is attached to the telescope (i.e. the ones at CASLEO and LCO). 
The spectra were processed and calibrated using standard routines in IRAF\footnote{
IRAF is distributed by the National Optical Astronomy Observatories, which are operated by
the Association of Universities for Research in Astronomy, Inc., under cooperative
agreement with the National Science Foundation.}, except FEROS data, for which
we applied the standard reduction pipeline provided by ESO (which uses comparison lamps observed
before and after each night).

\begin{table}
\caption{Details of the spectroscopic data for HD~74194.}
\begin{tabular}{l l c c c}
\hline\hline\noalign{\smallskip}
 Instr. config.    & Observatory & R     & Sp. range& N$^{\rm a}$ \\
                 &           &       & [nm]    &\\
\noalign{\smallskip}\hline\noalign{\smallskip}
Echelle, 2.5-m    & LCO       & 45\,000 &  345-985 &23\\
REOSC, 2.15-m     & CASLEO    & 15\,000 &  360-610 &14\\
FEROS, 2.2-m      & La Silla  & 46\,000 &  357-921 &19\\
\noalign{\smallskip}\hline
\multicolumn{5}{l}{a: N is the number of spectra.}\\
\end{tabular}
\label{runs}
\end{table}

\section{The radial-velocity orbit}

We measured the central wavelengths of the He {\sc ii} $\lambda\lambda$4542, 4686, and 5411 
absorption lines in the 56 spectra of HD~74194 and determined their RVs 
by means of the {\sc splot} task in {\sc iraf}. We chose these three lines because they 
present very similar RVs in a given spectrum. Other lines, such as He {\sc i} $\lambda\lambda$4471 and 
5875, present profile variations that prevent appropriate RV measurements.

The RVs are clearly variable \citep[as indicated by the first observations, ][]{2006ATel..819....1B}.
We therefore searched for periodicities in the whole RV dataset, and using the code published by 
\citet{1980PASP...92..700M}, we obtained a most probable period P=9.543 days. 
A very similar result of 9.542 days was given by  
the Lomb-Scargle algorithm \citep{1982ApJ...263..835S} provided online as a NASA Exoplanet Archive 
service \citep{2013PASP..125..989A}.

We determined the RV orbital solution by means of the {\sc gbart} code\footnote{{\sc gbart} is an
improved version of the original program for determining the orbital elements for 
spectroscopic binaries \citep{1969RA......8....1B} developed by F. Bareilles and available at http://www.iar.unlp.edu.ar/\textasciitilde fede/pub/gbart.}.
This code converges to a highly eccentric orbit, e=0.63$\pm$0.04 and  P=9.5436$\pm$0.0002 d.
We checked the {\sc gbart} orbital solution by means of the {\sc fotel} code developed by 
\citet{hadrava}, obtaining the same solution. The orbital elements determined by both methods 
are shown in Table~\ref{elements}. 
Thus, the new ephemeris of the system, in Heliocentric Julian days,  is
\begin{equation} \label{ephemeris}
 T_{\rm periastron}= 2\,454\,654.04 + 9.5436\times E
.\end{equation}

 \begin{table}[!t]
\caption{Orbital elements of HD~74194 determined by two independent methods.}
\label{elements}
\centering

\begin{tabular}{l c c}
\hline\hline\noalign{\smallskip}
Parameter&  {\sc gbart} & {\sc fotel}\\
\hline   \noalign{\smallskip}
$P$ [d]                             & 9.5436$\pm$0.0002  & 9.5436$\pm$0.0002\\
$e$                                 & 0.63$\pm$0.03      & 0.63$\pm$0.05    \\
$T_{\rm periastron}$ [HJD]$^a$    &634.95$\pm$0.04     & 634.95$\pm$0.05  \\
$T_{\rm VRmax}$ [km~s$^{-1}$]     &635.27$\pm$0.04     & 635.27$\pm$0.05  \\
$\omega$ [$^\circ$]               &302$\pm$4           & 302$\pm$4        \\
$V_{0}$ [km~s$^{-1}$]             &15.3$\pm$0.4        & 15.3$\pm$0.5       \\
$K$ [km~s$^{-1}$]                   &21$\pm$1            & 20.8$\pm$1.4     \\
$a \sin i$ [$R_\odot$]            &3.03$\pm$0.24       & 3.1$\pm$0.3      \\
$F(M)$ [$M_\odot$]                  &0.004$\pm$0.001     & 0.0042$\pm$0.0015\\
r.m.s. [km~s$^{-1}$]              &2.3                 &3.4               \\
\hline\noalign{\smallskip}
\multicolumn{3}{l}{a: Heliocentric Julian Day$-2\,454\,000$.}\\
\end{tabular}
\end{table}
   
   \begin{figure}
   \centering
   \includegraphics[width=9cm]{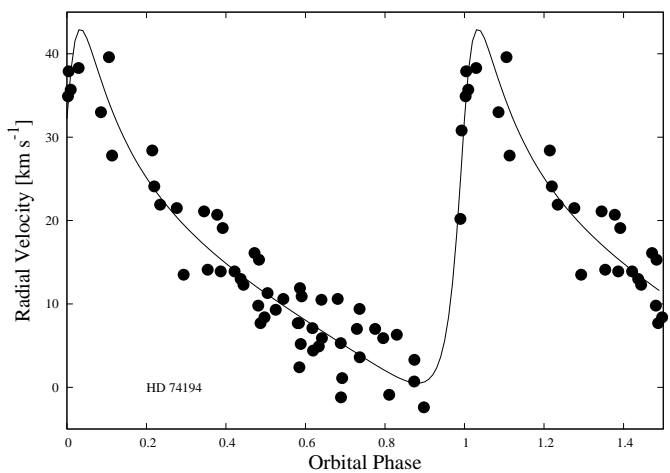}

      \caption{Radial-velocity orbit of HD~74194 derived from He {\sc ii} lines.}
         \label{sb1}
   \end{figure}
%

In addition,
we took advantage of the {\sc fotel} code to explore the possibility that this high-eccentricity system might
display apsidal motion. We tried solutions with the rate of apsidal advance ($\dot{\omega}$) as a free parameter,
and also split the RVs into two datasets covering the years 2006-2009 and 2010-2015.
Neither attempt gave definitive conclusions. This kind of analysis should be done with future observations with our solution as the initial epoch.
 
\section{Variability of the H$\alpha$ line}

We collected 42 spectra that cover the H$\alpha$ line.
The line is variable on a timescale of days in both intensity and the RV of the
absorption and emission components.
In most of the spectra, the H$\alpha$ line exhibits P Cygni-like profiles, with
an additional extended emission in the blue wing (described as
a {\em \emph{blue hump}} or Beal's Type III P Cygni profile).
In a few spectra, the absorption is located near the centre of the emission
profile, producing an apparent double emission with two peaks with similar
intensities.

We measured the RVs of the absorption and emission components in each
spectrum, whenever possible. 
The RVs of the emission peaks folded with the orbital ephemeris of equation (1)
roughly follow the same trend as the RVs of the He {\sc ii} absorption lines, although they show a larger dispersion with a semi-amplitude of $K=50\pm10$~km\,s$^{-1}$.
This behaviour suggests that the H$\alpha$ emission is associated with the O-type star.
On the other hand, the RVs of the absorption components do not follow the orbital period, and no other 
reliable periodicity was found (via the {\sc marmuz} program or the NASA Exoplanet Archive Periodogram 
Service). The material producing this absorption is not bound to the O-type star or to the secondary star.

\begin{figure}
   \centering
   \includegraphics[width=9cm]{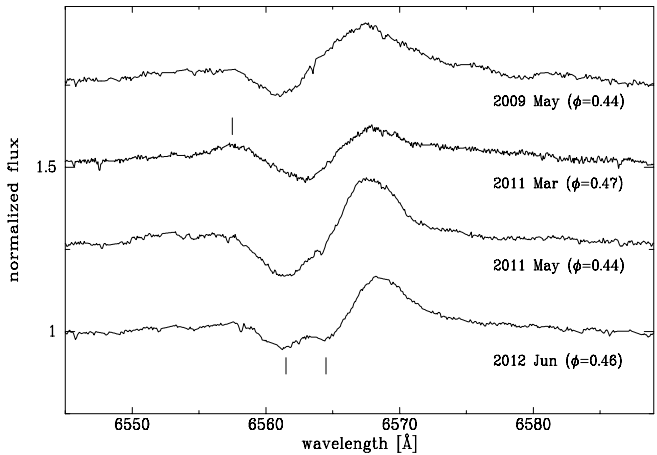}
      \caption{Variability of H$\alpha$ in the spectrum of HD~74194 at similar orbital phases.
              }
         \label{ha1}
   \end{figure}

We investigated whether the morphological variability of the complex H$\alpha$ line is related to the 
orbital period by comparing spectra taken at the same orbital phases.
For example, in Figure~\ref{ha1}, we compare four H$\alpha$ spectra taken at different epochs but similar 
orbital phases. The spectrum obtained in 2012 June exhibits a P-Cygni profile with a double
absorption component, whereas in the other spectra, the P-Cyg profiles present a
single absorption component.
Also, the extended emission in the blue wing (\emph{{\em blue hump}}) of the 2011 March spectrum is notable. 
Thus, the profile variability in H$\alpha$ is not entirely modulated with the orbital period. 

We also analysed the variability in the equivalent widths ($EW$) of the complex H$\alpha$ line. 
Such measurements are problematic because it is difficult to define the continuum in the echelle orders. 
We used the continuum of the next bluer order, which has no conspicuous spectral lines, and divided the order 
including H$\alpha$ by it (pixel to pixel). The spectra were also corrected by their respective RVs of 
the H$\alpha$ emission. Then, we determined the $EW$ in the resulting spectra, using the {\sc splot} task in 
IRAF in script mode to fix the wavelength limits between 6535\AA\ and 6580\AA\  
(thus including absorption and emission components). We searched for periodicities in the obtained 
$EWs$ with the 
NASA service. We obtained a most probable period of $P$=285.3 d, which is nearly 30 times the RV period.
The period uncertainty can be estimated as the full width at half maximum
of the periodogram peak (relative to power values); in this case, the result is about ten days.

%

   \begin{figure}
   \centering
   \includegraphics[width=9cm]{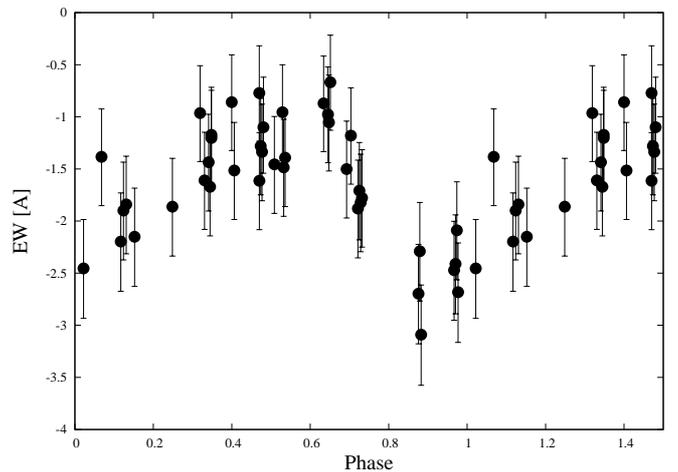}
      \caption{Variation in the $EW$ of H$\alpha$, phased with a period of 285.3 days and the time of periastron passage of the RV orbit. Bars represent the errors in the $EWs$. 
              }
         \label{ew}
   \end{figure}


\section{Discussion and conclusions}

We present for the first time the orbital solution of HD~74194, based on the RVs of optical absorption lines of He {\sc ii} measured in 56 spectra secured between 2006 and 2015.
The resulting orbit displays a high eccentricity ($e=0.63\pm0.03$) and a rather short periodicity ($P$=9.5436 d).
Such a combination of parameters is well outside the known distribution, 
in which there is no binary system with $P<10$ days and $e>0.43$, the most extreme previously known system being HD~37737 
\citep[$P=7.84$ d and $e=0.43$][]{2007ApJ...655..473M}.
Such close and highly eccentric binary orbits are thought to be possible in post-SN systems \citep{1996ApJ...471..352K}. 

\begin{figure}
   \centering
   \includegraphics[width=5.5cm, angle=270]{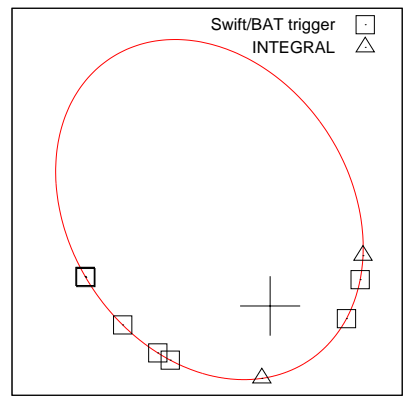}
      \caption{Positions, relative to the barycentre, of the O-type supergiant during the 
outbursts of HD~74194 found in the literature. 
              }
         \label{flares}
   \end{figure}

The new orbital ephemeris for HD~74194 supports analysis of the reported high-energy outbursts according to their phases.
We summarised these events in the Table~\ref{bursts}, where their respective orbital phases
(according to the ephemeris of Eq.~\ref{ephemeris}) are also shown.
It is remarkable that they occur in a restricted range of orbital phases, namely $\phi$=0.84--0.07, which is 
very near to the times of the periastron passage ($\phi$=0.00).
The apparent correlation between flares and periastron passages (see Fig.~\ref{flares}) 
seems to agree with the scenario of a compact star accreting matter from the clumpy wind of the supergiant.  
\citet{2010A&A...523A..90K} proposes different scenarios for this phenomenon, relating $P$ and $e$. 
HD~74194 falls into the SFXT regime, but very near to the unstable orbit region, where stars could evolve 
towards coalescence.

\onltab{
 \begin{table*}[!t]
\caption{Outbursts of HD~74194 found in the literature.}
\label{bursts}
\centering

\begin{tabular}{c l c c c c}
\hline\hline\noalign{\smallskip}
Id. &Date&  Cite & Mission& HJD &$\phi$ \\
\hline   \noalign{\smallskip}
1&2003-07-01 & \citet{2007ApJ...655L.101G} &{\it INTEGRAL}& $2\,452\,822.333$ &0.07\\
2&2006-05-15 & \citet{2007ApJ...655L.101G} &{\it INTEGRAL}& $2\,453\,871.271$ &0.98\\
3&2006-10-04 & \citet{2007ApJ...655L.101G} &$Swift$       & $2\,454\,013.115$ &0.84\\
4&2008-07-05 & \citet{2009MNRAS.392...45R} &$Swift$       & $2\,454\,653.385$ &0.93\\
5&2008-09-21 & \citet{2009MNRAS.397.1528S} &$Swift$       & $2\,454\,730.828$ &0.05\\
6&2009-08-28 & \citet{2009ATel.2178....1B} &$Swift$       & $2\,455\,072.453$ &0.84\\
7&2010-03-28 & \citet{2010ATel.2520....1R} &$Swift$       & $2\,455\,284.162$ &0.03\\
8&2011-08-25 & \citet{2011ATel.3586....1M} &$Swift$       & $2\,455\,798.537$ &0.92\\
9&2013-07-02 & \citet{2013ATel.5190....1R} &$Swift$       & $2\,456\,475.841$ &0.89\\
\hline
\end{tabular}
\end{table*}
}

%

\begin{figure}
   \centering
   \includegraphics[width=6.0cm, angle=270]{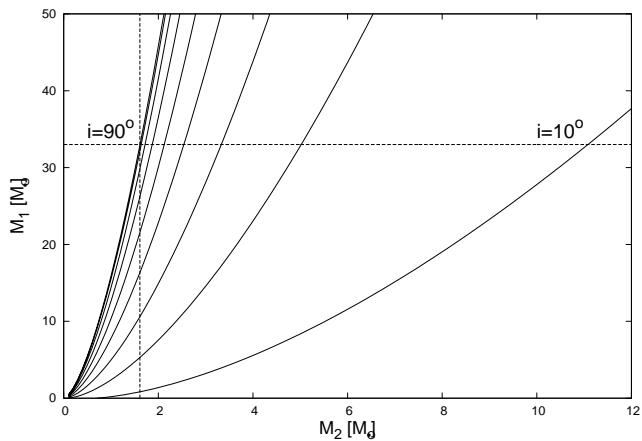}
      \caption{Mass of the primary as a function of the secondary mass, derived by fixing the mass 
function to the
value of Table~\ref{elements}. Curves were calculated for each orbital inclination from 90$^\circ$ to 
10$^\circ$ in steps of 10$^\circ$. Dotted lines indicate the primary mass, M$_1$=33 M$_\odot$,
and the minimum possible secondary mass, M$_2$=1.61 M$_\odot$, for reference. 
              }
         \label{masses}
   \end{figure}

We can discuss the unknown secondary component further. We did not find any traces of spectral features belonging to the companion.
Since we have only determined the mass function ($F(M)$, in Table~\ref{elements}), which relates both
masses and the orbital inclination, it is not possible to determine the secondary
(minimum) mass unequivocally, but some analysis can be done.
In Fig.~\ref{masses}, we depict how the primary mass (M$_1$) varies as a function of the secondary
mass (M$_2$) and the orbital inclination($i$).
If we assume that the mass of the primary O8.5 Ib-II star is 33 M$_\odot$ \citep[average of both calibrations
in][]{2005A&A...436.1049M},
it can be inferred that no secondary mass lower than 1.61 M$_\odot$ is possible. 
This limit is near the high end of the neutron-star (NS) range. 
In a recent work, \citet{2013ApJ...778...66K} discuss the NS mass distribution. 
They find that NS masses peak at 1.33 M$_\odot$ and 1.55 M$_\odot$ depending whether they are in 
NS-NS or NS-white dwarf (WD) binary systems, respectively. \citet{2012ApJ...757...55O} 
also discuss the NS masses in other populations of binaries, obtaining a mean of 
1.28 M$_\odot$ and a dispersion of 0.24 M$_\odot$ in 
eclipsing binaries with high-mass primaries \citep[six values from][]{2011ApJ...730...25R}.
However, there are some NSs with higher values: 
Vela X-1 \citep[1.84$\pm$0.06 M$_\odot$,][]{2011ApJ...730...25R};
\object{PSR J1748-2446I} \citep[M=1.91+0.02-0.10 M$_\odot$,][]{2013ApJ...778...66K};
\object{PSR J1614-2230} \citep[M=1.97$\pm$0.04 M$_\odot$][]{2010Natur.467.1081D};
\object{PSR B1516+02B} \citep[M=2.10$\pm$0.19 M$_\odot$,][]{2013ApJ...778...66K}; 
and possibly the ``Black Widow'' pulsar 
\object{PSR B1957+20} with M=2.40$\pm$0.12M$_\odot$ \citep{2011ApJ...728...95V}.

\begin{figure}
   \centering
   \includegraphics[width=6.0cm, angle=270]{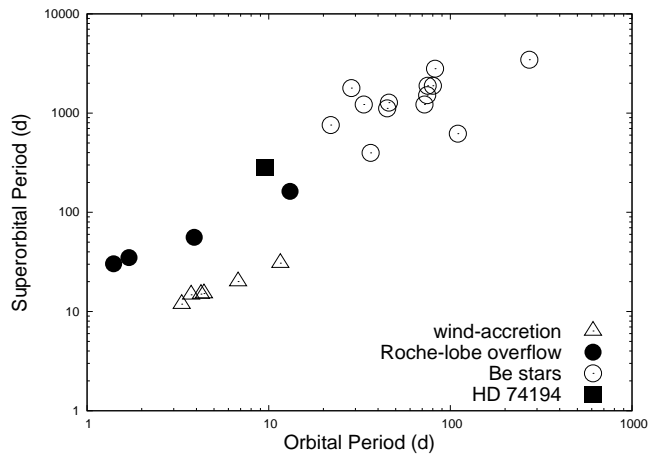}
      \caption{Position of HD~74194 in the superorbital- vs. orbital-period plot for wind accretion, 
Roche-lobe overflow systems, and Be stars taken from \citet{2013ApJ...778...45C} and 
\citet{2011MNRAS.413.1600R}.   
              }
         \label{superorbital}
   \end{figure}

Since the H$\alpha$ line varies but not in relation to the orbital period, 
we looked for superorbital modulation as found in other similar binary systems.
We found a periodicity of about 285 d in the $EW$ (considering the absorption and emission components
as a whole), but it should be confirmed by further observations.
A relationship between superorbital and orbital periods was discussed by 
\citet{2013ApJ...778...45C} for a wide variety of X-ray binaries.
Our preliminary superorbital period is nearly 30 times the orbital one, which is very far from
the relation that seems to exist for the wind-accretion high-mass X-ray binaries, but closer to the one for 
Roche-lobe overflow powered systems (see Fig.~\ref{superorbital}).

%

\begin{acknowledgements}
We thank the referee for comments that helped to improve the paper.
We thank the directors and staffs of CASLEO, LCO, and ESO La Silla for the use of their facilities and their kind hospitality during the observing runs.
CASLEO is operated under agreement between CONICET and the 
Universities of La Plata, C\'ordoba and San Juan, Argentina.
RHB acknowledges FONDECYT Project No. 1140076.
JIA aknowledges financial support from FONDECYT Project No. 11121550.
JMA, AS and EJA acknowledge support from the Spanish Ministry for
Economy and Competitiveness and FEDER funds through grant AYA2013-40611-P.
This research made use of NASA's Astrophysics Data System and the SIMBAD database, operated at the CDS, Strasbourg, France.
\end{acknowledgements}

\bibliographystyle{aa}
\bibliography{own}
\end{document}